\documentclass[preprintnumbers,amsmath,amssymb,groupedaddresst,nofootinbib]{revtex4}
\usepackage{graphicx}

\usepackage{color}
\usepackage{bm}
\usepackage{subfigure}

\begin{document}

\preprint{AP-GR-118, OCU-PHYS-416, OU-HET-840, RIKEN-MP-98}

\title{Meson turbulence at quark deconfinement from AdS/CFT}

\author{Koji Hashimoto$^{1,2}$}
\author{Shunichiro Kinoshita$^{3}$}
\author{Keiju Murata$^{4}$}
\author{Takashi Oka$^{5}$}
\affiliation{$^{1}${\it Department of Physics, Osaka University,
Toyonaka, Osaka 560-0043, Japan}}
\affiliation{$^{2}${\it Mathematical Physics Lab., RIKEN Nishina Center,
Saitama 351-0198, Japan}}
\affiliation{$^{3}${\it Osaka City University Advanced Mathematical Institute, Osaka 558-8585, Japan}}
\affiliation{$^{4}${\it Keio University, 4-1-1 Hiyoshi, Yokohama 223-8521, Japan}}
\affiliation{$^{5}${\it Department of Applied Physics, University of Tokyo, 
Tokyo 113-8656, Japan}}

\begin{abstract}
Based on the QCD string picture at confining phase, we conjecture that the 
deconfinement transition always accompanies a condensation
of higher meson resonances with a power-law behavior, ``meson turbulence''.
We employ the AdS/CFT correspondence to calculate the meson turbulence for
${\cal N}=2$ supersymmetric QCD at large $N_c$ and at strong coupling limit,
and find that the energy distribution to each meson level $n$ scales as $n^\alpha$
with the universal scaling $\alpha=-5$. The universality is checked for various
ways to attain the quark deconfinement: a static electric field below/around the critical value,
a time-dependent electric field quench, and a time-dependent quark mass quench,
all result in the turbulent meson condensation with the 
universal power $\alpha=-5$ around the deconfinement.
\end{abstract}

\maketitle


\section{Turbulence and quark deconfinement}

How the quarks are confined at the vacuum of quantum chromodynamics (QCD) is
one of the most fundamental questions in the standard model of particle physics.
The question has attracted attention for long years, and recently investigation has
diverse approaches. The question is difficult simply because of the fact that the confinement appears at the vacuum, not in a particular corner with specific external forces.
Therefore, the confining vacuum can be broken in various manner as one departs from
the vacuum with the help of some external forces. The forces include for example a finite temperature, a finite quark density and electric fields. Depending on how you break the vacuum confinement, the resultant deconfined phases show various aspects with various global symmetries. This variety makes the confinement problem even more difficult to be understood. 

\vspace{5pt}

We would like to find a universal feature of the deconfinement. To understand the nature of the quark confinement, we need a proper observable which exhibits a universal behavior irrespective of how we break the confinement. In this paper, we propose  
a universal behavior of resonant mesons and name it  {\it meson turbulence}.

\vspace{5pt}

As we have summarized in our letter \cite{Previous}, 
a particular behavior of resonant mesons (excited states of mesons) can be an indicator of the deconfinement. The meson turbulence 
is a power-law scaling of the resonant meson condensations. For the the
resonant meson level $n$ ($n=0,1,2,\cdots$), the condensation of the
meson $\langle c_n(x,t)\rangle$ with its mass $\omega_n$ causes
the $n$-th meson energy $\varepsilon_n$ scaling as
$(\omega_n)^\alpha$ with a constant power $\alpha$.  
This coefficient $\alpha$ will be unique for a given theory, and 
does not depend on how one breaks the confinement. 
In particular, for the theory which we analyze in this paper, that is ${\cal N}=2$ supersymmetric QCD with ${\cal N}=4$ supersymmetric Yang-Mills as its gluon sector at large $N_c$ at strong coupling, the universal power-law scaling parameter $\alpha$ is found to be 
\begin{eqnarray}
\langle \varepsilon_n \rangle \propto (\omega_n)^{\alpha}, \quad \alpha = -5 \, .
\end{eqnarray}
where $\varepsilon_n$ is the energy of the $n$-th meson resonance.
Normally, for example at a finite temperature, the energy stored at the $n$-th level of the 
resonant meson should be a thermal distribution, $\varepsilon_n\propto \exp[-\omega_n/T]$.
The thermal distribution is Maxwell-Boltzmann statistics, in which the higher (more massive) meson modes are exponentially suppressed. 
However, we conjecture that this standard exponential suppression 
will be replaced by a power-law near any kind of the deconfinement transitions.
If we think of the meson resonant level $n$ as a kind of internal momentum,
then the energy flow to higher $n$ can be regarded as a so-called 
weak turbulence. This
is why we call the phenomenon meson turbulence, and the level
$n$ can be indeed regarded as a momentum in holographic direction in the AdS/CFT correspondence \cite{Maldacena:1997re,Gubser:1998bc,Witten:1998qj}.

\vspace{5pt}

\begin{figure}
\includegraphics[width=10cm]{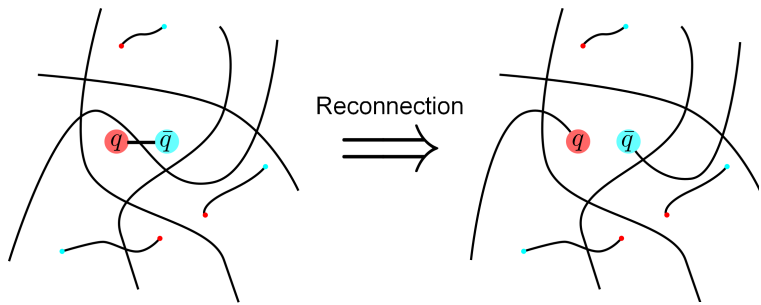}
\caption{A schematic picture of the deconfinement phase as condensation of QCD strings.
Left: we add a meson (a pair of a quark and an anti-quark connected by a QCD string) to the system.
Right: due to the background condensed QCD strings, the QCD string can be reconnected, and the 
quark can freely propagate away from the anti-quark.}
\label{figreconnect}
\end{figure}

The reason we came to the universal power behavior is quite simple. We combined two
well-known things,
\begin{itemize}
\item Mesons are excitations of an open  QCD string.

As is well-known, mesons and their resonant spectra are described by a quark model with
a confining potential. The confining potential has a physical picture of an open string whose end points are quarks. Rotating strings can reproduce Regge behavior of the meson resonant spectra.

\item Deconfinement phase is described by a condensation of long strings.

It has been argued that the deconfinement phase can be identified as a condensation of
long QCD strings \cite{Polyakov:1978vu} 
(see \cite{Lucini:2005vg,Hanada:2014noa,Pisarski:1982cn,Patel:1983sc}). Once long QCD
strings are condensed in the background, if one adds a quark antiquark pair to that, the
string connecting the quark antiquark pair can be reconnected with the background long QCD strings condensed, then the quark can propagate freely away from the antiquark.  (See Fig.~\ref{figreconnect}).
So the presence of the background QCD strings realizes the deconfinement.
The picture is familiar in view of the renowned 
dual Meissner effect. Superconducting phase can be broken in a large magnetic field: many magnetic vortex strings are produced, and normal phase may be understood as condensed vortex strings. Upon the duality, the vortex strings correspond to the QCD strings, and confining (deconfining) phase corresponds to
superconducting (normal) phase. 

\end{itemize}
Combining these two leads us to the conjecture that {\it the deconfinement of quarks is indicated by a condensation of higher meson resonances}. More precisely, we
claim that {\it the condensation should be turbulent}: the higher mode condensation is not suppressed exponentially 
but behaves with a power-law.

\vspace{5pt}

The reason we expect it to be a power law scaling is intuitively from a Hagedorn transition in string theory. String theory was born as an effective theory of hadrons at low energy, and is believed to be a good approximation at a large $N_c$ limit of QCD. Since the number of states in string theory grows exponentially, a free string theory can reach an upper bound of
the temperature which is called Hagedorn temperature. The exponential suppression of
the higher states (Maxwell-Boltzmann statistics) is canceled by the growth of the number of states, and near the critical temperature the energy distribution changes from the exponential suppression law 
to a power-law.

\vspace{5pt}

The energy flow from larger scales to smaller scales with a universal
power-law reminds us of turbulence in fluid dynamics. 
Indeed, 
the energy transfer between different scales because of
non-linearities in some non-linear systems is called weak turbulence, 
and the Kolmogorov scaling factor $-5/3$ is universal in fluids. The name ``meson turbulence'' simply means that the energy distribution of the meson modes as a function of the meson resonant level $n$
is the power-law with a universal power $\alpha$. Here we regard the meson resonant level $n$ as a kind of momentum, such that a larger $n$ means a smaller scale. 
We are also motivated from a recent discovery of a weak turbulence in gravity in AdS, a Bizon-Rostworowski conjecture\cite{Bizon:2011gg}. There, a power-law scaling of the energy distribution in the AdS gravity was found and is found to be universal for various initial conditions, indicating a turbulence.
Since we shall work in the gauge/gravity 
correspondence, our meson turbulence may serve as 
a quark counterpart of the AdS instability.

\vspace{5pt}

In QCD, the phase transition region in the phase diagram is thought to be still at a strongly coupled regime, we in this paper employ the renowned AdS/CFT correspondence
\cite{Maldacena:1997re,Gubser:1998bc,Witten:1998qj}. Since anyhow the AdS/CFT correspondence is expected to work well only at a large $N_c$ limit which is different from 
the realistic QCD, we use the most popular example in the AdS/CFT
analysis: the ${\cal N}=2$ supersymmetric QCD 
realized by the D3/D7 brane system~\cite{Karch}. 
Fortunately, the theory has a discrete meson states
although the gluon sector is the ${\cal N}=4$ supersymmetric Yang-Mills theory and thus is deconfined. We can concentrate on quark deconfinement, not the gluon deconfinement, in the model. The situation is somewhat similar to the heavy ion collision experiments as heavy quarkonia can survive even in the quark gluon plasma.
In the AdS/CFT correspondence, there appears infinite number of meson resonances.
They are just Fourier modes in holographic dimensions. So, higher meson resonant modes labeled by a large $n$ corresponds to a smaller scale in the gravity dual,
so our name ``meson turbulence'' can make sense in the gravity dual as a weak turbulence for the energy flow into a high frequency modes in extra holographic dimensions.

The meson turbulence is not the turbulence in the real space but in the holographic space, thus the power law is
not of the real space momentum $p$ but of the meson level $n$. The real-space turbulence in QCD has been
studied in relation to thermalization at heavy ion collisions \cite{Arnold:2005ef,Arnold:2005qs,Mueller:2006up,Arnold:2007cg,Rebhan:2008uj,Berges:2008mr,Dusling:2010rm,Carrington:2010sz,Blaizot:2011xf,Florchinger:2011qf,Fukushima:2011ca,Fukushima:2013dma}, based on general grounds for scalar/gauge field theories \cite{Micha:2002ey,Micha:2004bv,Berges:2008wm,Berges:2008sr,Berges:2010ez,Berges:2012us,Berges:2012ev,Schlichting:2012es,Berges:2013fga,Berges:2008zt}.
Note the difference from our turbulence in the holographic space.

\vspace{5pt}

We shall investigate various deconfining transitions in this paper, to check the universality of our conjecture of the meson turbulence. First, we work with a static case. A nonzero electric field is a good example since a strong electric field can make the quark-antiquark pair dissociate. Beyond the value of the electric field called Schwinger limit, the system experiences a phase transition from the insulator (confining) phase to a deconfined phase with electric quark current flow. We look at the situation just before/around 
the phase transition, and will find the meson turbulence.

\vspace{5pt}

Then we investigate time-dependent setup. The virtue of the AdS/CFT correspondence is 
that time-dependent analysis is possible, as opposed to lattice simulations of QCD. To demonstrate the universality of the meson turbulence, we work in two examples: 
(1) electric-field quench, and (2) quark-mass quench. In (1), we change the electric field in a time-dependent manner, from zero to a nonzero finite value, 
in a duration denoted by $\Delta t$. Then the system dynamically evolves and we follow it in the gravity dual of the gauge theory. At later times a singularity is formed, which signals the deconfinement \cite{Hashimoto:2014yza}. 
We perform a spectral analysis of the mesons and find that, at high frequencies, the system 
is transferred from an initial spectrum determined by the quench
to the power-law, with the universal coefficient $\alpha = -5$. Then in (2), we consider a time-dependent change in quark mass.
The quark mass starts to grow and then comes back to the original value,
in the duration $\Delta t$. 
In this case we will not include the worldvolume gauge field but excite
only the brane fluctuations, because we attempt to explore the essence of
the meson turbulence in the simplest setup. 
We again find the meson turbulence and the universal power law with the power $\alpha = -5$.

\vspace{5pt}

The universality we found in this paper strongly indicates that the meson turbulence is 
a universal phenomena which is independent of how one breaks the confinement. The condensation of long QCD strings in the completely deconfined phase is a difficult task, 
so we look into the behavior of mesons just before the deconfinement. We hope that our approach would serve as a new approach to understand the quark confinement problem.

\vspace{5pt}

The organization of this paper is as follows. First, we shall give a brief review of the 
holographic setup to introduce our notation, in particular the meson
effective action given by the AdS/CFT correspondence. Then in section~\ref{sec:static},
we provide an analysis for the static case with the electric field. We
illustrate the example with an evaluation of a vanishing string
tension. In section~\ref{sec:electric_quench}, we recall quenches in linear theory and examine non-linear evolutions of the electric field quench. 
In section~\ref{sec:mass_quench}, 
we analyze the quark-mass quench as the simplest setup and demonstrate the universality of the meson turbulence.
The final section is devoted to a summary.

\section{Review: ${\cal N}=2$ supersymmetric QCD in holography}

The simplest set-up in string theory which accommodates quarks in four spacetime dimensions 
is the ${\cal N}=2$ supersymmetric QCD. More precisely, the flavor quark sector is added as
a kind of a probe and so the number of flavor is $N_{\rm f}=1$, while the gauge group is
$U(N_c)$ with a large $N_c$ limit, $N_c \rightarrow \infty$. The gluon part has the maximal 
supersymmetries, ${\cal N}=4$. This set-up served as a best toy model of QCD in the AdS/CFT correspondence \cite{Karch}. To obtain the gravity dual, we take the strong coupling limit $\lambda \equiv N_c g_{\rm YM}^2 \rightarrow \infty$, too. Then the correspondence states that the meson sector is nothing but the flavor D7-brane action in the $\mathrm{AdS}_5\times S^5$ background geometry,
\begin{eqnarray}
&&
S =  \frac{-1}{(2\pi)^6 g_{\rm YM}^2 \l_s^8}\int \!\! d^8\xi \sqrt{
-\det (g_{ab} [w]+ 2\pi l_s^2 F_{ab})} \, ,
\label{D7action}
\\
&&
ds^2 = \frac{r^2}{R^2}\eta_{\mu\nu}dx^\mu dx^\nu \!+\! \frac{R^2}{r^2}
\!\left[
d\rho^2 \!+\! \rho^2 d\Omega_3^2 \!+\! dw^2\!+\!d\bar{w}^2
\right]\, ,
\nonumber
\end{eqnarray}
where $r^2 \equiv \rho^2 + w^2 + \bar{w}^2$, 
$F_{ab}=\partial_a A_b-\partial_b A_a$, 
and $R\equiv (2\lambda)^{1/4} l_s$ is the $\mathrm{AdS}_5$ curvature radius.
The fields on the D7-brane, $A_a$ and $w$, are on ($7+1$)-dimensional worldvolume.

\vspace{5pt}

We assume a trivial dependence on the internal direction within $S^5$ since they are
expected to be irrelevant to the real QCD dynamics. Then the fields are dependent only on the AdS radial direction $\rho$ and our ($3+1$)-dimensional spacetime,
$w(x^\mu,\rho)$ and $A_a(x^\mu,\rho)$. Basically the decomposition of these five-dimensional 
fields into eigen components in the $\rho$ direction gives a Kaluza-Klein-like tower of infinite number of four-dimensional fields. The four-dimensional fields are the scalar and the vector mesons. So, the D7-brane action (\ref{D7action}) is nothing but the meson effective action.

\vspace{5pt}

Let us write the meson decomposition and the effective action explicitly.
A convenient rescaling for the vector field is
$a_a\equiv 2\pi l_s^2 R^{-2} A_a$.
Furthermore, we set $\bar{w}(x^\mu,\rho)=0$ for simplicity.
The theory has only a single scale which is the quark mass.
It is defined by the location of
the D7-brane at the asymptotic AdS boundary, 
as $w(x^\mu,\rho=\infty)=R^2 m$, where 
the quark mass $m_q$ is given by this $m$ as
$m_q=(\lambda/2\pi^2)^{1/2}m$.
The static solution of the shape of the D7-brane with this boundary condition at the 
AdS boundary is simply a straight D7-brane,
\begin{eqnarray}
w(x^\mu,\rho)=R^2m, \qquad 
a_a(x^\mu,\rho)=0.
\end{eqnarray}
We consider a fluctuation of the fields around this background solution, and find 
a fluctuation action explicitly written to the second order in the fluctuation, as \cite{Kruczenski:2003be}
\begin{equation}
S= \int \!dt d^3x\int^1_0 \!dz \frac{1-z^2}{2z}[
(\partial_t\bm{\chi})^2-m^2(1-z^2)(\partial_z\bm{\chi})^2
]\ + {\cal O} (\bm{\chi}^3),
\nonumber
\end{equation}
where we have defined the fluctuation fields as
$\bm{\chi}\equiv (R^{-2}w-m,a_x)$, and assumed that all fields are independent of
the spatial coordinates $x^1,x^2,x^3$. 
Note that $a_x$ denotes the vector field along an spatially-homogeneous
direction in the Euclidean space with these coordinates.
An irrelevant overall factor is neglected in the action. 
The new radial coordinate $z$ is defined through
\begin{eqnarray}
\rho\equiv R^2m\frac{\sqrt{1-z^2}}{z}\, .
\end{eqnarray}
In this new coordinate, 
$z=0$ is the AdS boundary, and $z=1$ is the D7-brane center that is closest to
the Poincar\'e horizon in the bulk AdS.

\vspace{5pt}

To calculate the meson modes, we solve 
the equation of motion for $\bm{\chi}$ to the leading order in the fluctuation, 
\begin{equation}
 \left[\frac{\partial^2}{\partial t^2} 
 -
 m^2
 \frac{z}{1\!-\!z^2}
 \frac{\partial}{\partial z}\frac{(1\!-\!z^2)^2}{z}
 \frac{\partial}{\partial z}
\right]\bm{\chi}=0\ .
\label{lineq}
\end{equation}
This can be solved with 
\begin{eqnarray}
&& \chi = \sum_{n=0}^\infty \textrm{Re}\left[C_n \exp[i\omega_n t] e_n(z)\right],
\\
&& e_n(z)\equiv \sqrt{2(2n+3)(n+1)(n+2)} \, z^2 F(n+3,-n,2;z^2)\ ,
\label{eq:normal_mode}
\end{eqnarray}
where
\begin{eqnarray}
\omega_n \equiv 2\sqrt{(n+1)(n+2)}\, m \label{eq:meson_mass}
\end{eqnarray}
 is the resonant meson mass for the meson level number $n=0,1,2,\cdots$,
and  $F$ is the Gaussian hypergeometric function. Note that the resonant meson mass is almost equally spaced but not exactly. 

\vspace{5pt}

The  inner product in the $z$-space is defined as
\begin{eqnarray}
(f,g)\equiv \int^1_0 dz\, z^{-1}(1-z^2)f(z)g(z)\ ,
\end{eqnarray}
with which the eigen mode functions satisfy the ortho-normality condition
\begin{eqnarray}
(e_n,e_m)=\delta_{mn} \, .
\label{ortho}
\end{eqnarray}
In particular, the eigen functions $e_n(z)$ are normalizable. 
Note that an external electric field with $a_x=-Et$ 
satisfies Eq.~(\ref{lineq}) and it is non-normalizable, so giving a constant electric field background. 
Using the eigen modes, to derive the meson effective action we expand 
the scalar field and the vector field as 
\begin{eqnarray}
\bm{\chi}=(0,-Et)+\sum_{n=0}^\infty \bm{c}_n(t)e_n(z)\, .
\label{decompomeson}
\end{eqnarray}
In this expansion, the coefficient fields
$\bm{c}_n(t)$ are meson fields, which share the same quantum charge. 
The non-negative integer $n$ denotes the resonant level of the meson.
The meson field $\bm{c}_n(t)$ corresponds to some quark bilinear operators such
as $\bar{\psi}\psi$ and $\bar{\psi}\gamma_\mu \psi$, but there are important differences:
First, the meson fields are fluctuation around the hadronic vacuum. So the value 
of the meson field, which is zero at the hadron vacuum, 
is not directly related to the operator expression, for example, $\langle\bar{\psi}\psi\rangle$
which is nonzero even at the hadron vacuum. Second, the normalization of the meson
fields is determined such that the effective action of the meson has a canonically normalized
kinetic term, so the relation to the operator expression is up to a normalization factor.

\vspace{5pt}

Substituting
Eq.~(\ref{decompomeson})
back to Eq.~(\ref{D7action}), we obtain the meson effective action
\begin{eqnarray}
S = \frac12 \int \! d^4x\;
\sum_{n=0}^\infty \left[\dot{\bm{c}}_n^2 - \omega_n^2 \bm{c}_n^2 \right] + \mbox{interaction} \, ,
\label{mesoneff}
\end{eqnarray}
where we have omitted a constant term and total derivative terms.
One can work out all the nonlinear interaction terms if one wishes.
We can define the energy stored in the $n$-th meson resonance as
\begin{eqnarray}
\varepsilon_n\equiv \frac{1}{2}(\dot{\bm{c}}_n^2+\omega_n^2 \bm{c}_n^2)
\label{energyn}
\end{eqnarray}
and also the linearized total energy 
\begin{eqnarray}
\varepsilon=\sum_{n=0}^{\infty} \varepsilon_n\, .
\end{eqnarray}
These expressions for the meson energies will be used later for evaluating how much energy
is stored to the level $n$ meson resonance.

\vspace{5pt}

The important term is the interaction terms in the meson effective action. They make an energy transfer from one level to another. We need not to explicitly calculate the interaction terms in terms of the component meson resonant fields. In the following sections, we use the
D7-brane action itself to calculate the scalar/vector fields in a time-dependent manner,
and then decompose the fields into the meson resonances by using the inner-product and the ortho-normality (\ref{ortho}).


\section{Meson turbulence at static electric field}
\label{sec:static}

\subsection{Bending of the D7-brane}

In this section, we shall evaluate the turbulent meson condensation for a deconfinement caused by a static 
and constant electric field. We first solve the shape of the D7-brane for the finite electric field, and decompose 
the scalar field into the meson resonance. 

\vspace{5pt}

\begin{figure}
\includegraphics[width=6.4cm]{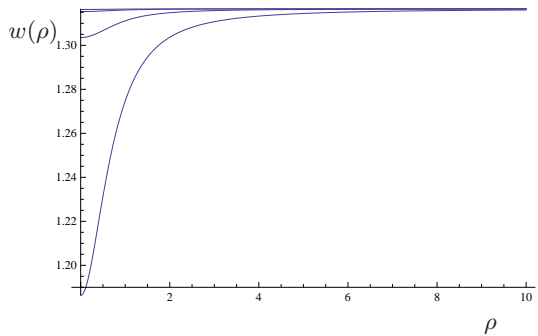}
\put(-20,-10){$\rho$}
\put(-200,100){$w(\rho)$}
\caption{The shape of the D7-brane in the AdS Schwarzschild background. 
As the temperature of the background geometry increases,
the shape changes and the D7-brane bends toward the black hole.}
\label{figTshape}
\end{figure}

But before considering the electric field, it would be instructive to consider the case with a finite temperature,
as a warm-up example to grasp what will happen to the probe D7-brane around the deconfinement transition.
Once we turn on a temperature, the background geometry of AdS${}_5$ starts to deform
and has a horizon. The D7-brane starts to be deformed, see Fig.~\ref{figTshape}. This deformation is
due to a supersymmetry breaking by the black hole horizon (or by the temperature). The central part of the
D7-brane is attracted to the black hole horizon, and finally is absorbed into the horizon. The D7-brane before hitting
the black hole horizon is called ``Minkowski embedding'', while the D7-brane whose tip is already inside the horizon
is called ``black hole embedding''. When a part of the D7-brane is inside the horizon, the spectrum of the
fields on the D7-brane becomes continuous, which means that the mesons melt and we are at a deconfined phase.
So, hitting the black hole horizon is the deconfinement transition.

\vspace{5pt}

From Fig.~\ref{figTshape} we can see that the D7-brane gets sharper and sharper as we increase the
temperature. And in fact, at the deconfinement transition, the D7-brane needs to be conical if we assume that the
deformation is smooth. This is because at the black hole horizon the redshift factor is infinite, so anything should
be perpendicular to the horizon. In order to have this conical shape, we need to turn on higher resonant modes
of mesons, since the mode (resonant level) number is almost like a Fourier level. This basically tells us that the
deconfinement transition is accompanied with higher resonant meson condensation, that is, the turbulent meson condensation.

\vspace{5pt}

\begin{figure}
\includegraphics[width=6.5cm]{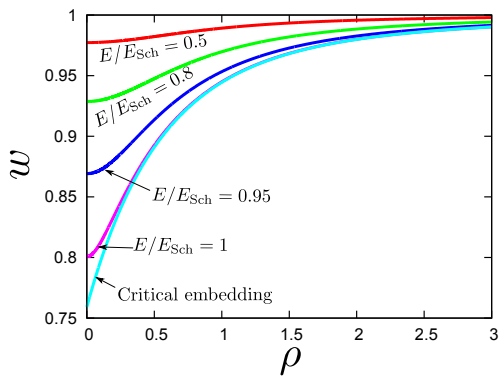}
\caption{The shape of the probe D7-brane in static electric fields
in the unit of $R=m=1$. 
The lines correspond respectively to
$E/E_\textrm{Sch}=0.5, 0.8, 0.95, 1$ and the critical embedding from top to bottom.
}
\label{figEshape}
\end{figure}

Now we proceed to the case with a finite electric field on the D7-brane. In fact, the situation is
quite similar to the case of the finite temperature black hole in the background. First, let us look at the
D7-brane shape with the finite electric field on the D7-brane, see Fig.~\ref{figEshape}. 
It is a plot of the solution of the equations of motion of the D7-brane action (\ref{D7action}) with
the electric field is introduced by a solution $a_x = -Et$. The calculation was originally performed in
Refs.~\cite{Karch:2007pd,Albash:2007bq,Erdmenger:2007bn}. 
The shape has a similar structure to the case of the finite temperature. As we increase the electric field $E$, 
the D7-brane bends toward the Poincar\'e horizon of the AdS geometry. 

\vspace{5pt}

\begin{figure}
\includegraphics[width=6.5cm]{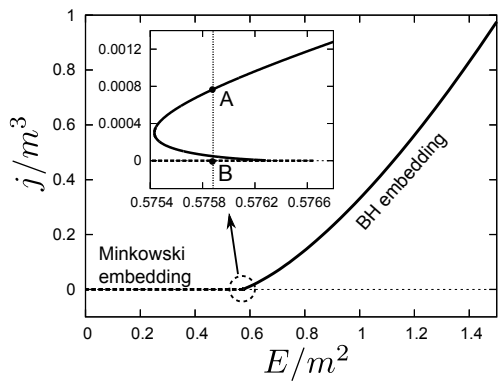}
\caption{The electric current $j$ as a function of the electric field $E$. At the line AB, the phase transition occurs
where $E=E_{\rm Sch}$. The transition is the formation of an effective horizon of the world volume of the D7-brane.
The D7-brane configuration exists even beyond $E=E_{\rm Sch}$, 
and the largest $E$ with the Minkowski embedding is called critical embedding which
is the phase boundary between the Minkowski and black hole embeddings.}
\label{Ej}
\end{figure}

Figure~\ref{figEshape} is with various value of the electric field in the unit of the 
Schwinger limit $E=E_\textrm{Sch}=0.5759m^2$ (in the unit $R=1$).
Beyond the Schwinger limit of the electric field, 
a first order phase transition to deconfinement occurs 
\cite{Albash:2007bq,Erdmenger:2007bn}.
In Fig.~\ref{figEshape}, we also plot the case with a critical embedding.
At the critical embedding, the bending of the D7-brane is the sharpest.
The critical embedding is a boundary of the Minkowski embedding and the
black hole embedding, and is not favored thermodynamically, see the phase diagram 
of the electric field $E$ versus the electric current $j$, Fig.~\ref{Ej}.
At point AB (the Schwinger limit) the phase transition occurs, but the Minkowski embedding is possible even beyond
$E=E_{\rm Sch}$. The end point to the right of the point B is the critical embedding.
The importance of the critical embedding is that it is a phase boundary and reflects the information of the black hole embedding most. At the critical embedding, the
brane shape is conical.

\vspace{5pt}

%

\subsection{Turbulent meson condensation}

Let us examine the meson condensation for each resonant mode.
We have learned that at the critical embedding which is the phase boundary to the black hole embedding
the D7-brane's shape is conical. Once the brane is conical, it needs a condensation of infinitely high Fourier modes.
This signals the turbulent meson condensation.

\vspace{5pt}

\begin{figure}
\includegraphics[width=6.7cm]{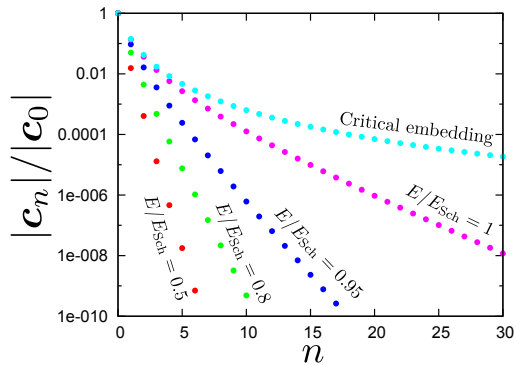}
\caption{Decomposed meson condensate 
$\log[|\bm{c}_n|/|\bm{c}_0|]$ in static electric fields. 
Colors Red/Green/Blue/Magenta/Cyan 
correspond respectively to $E/E_\textrm{Sch}=0.5, 0.8, 0.95, 1$
and the critical embedding.
}
\label{figEdecomp}
\end{figure}

To look at how each meson behaves as we approach the critical embedding,
we decompose the solution (shape) $\bm{\chi}(t,z)$ of the D7-brane action (\ref{D7action}) 
into the meson resonant modes, Eq.~(\ref{decompomeson}). 
The result is shown in Fig.~\ref{figEdecomp}. There we plot 
the ratio $|\bm{c}_n|/|\bm{c}_0|$ as a function of $n$ 
where we define 
$|\bm{c}_n| \equiv \sqrt{(c_n^{\rm scalar})^2+(c_n^{\rm vector})^2}$ for illustration.
We find two facts:
\begin{itemize}
\item
As we increase the electric field, the condensation of the high resonant modes
($\bm{c}_n, n\gg 1$)
get more enhanced.
\item At the critical embedding, the higher resonant modes participate stronger
than the Maxwell-Boltzmann law.
\end{itemize}

\vspace{5pt}

\begin{figure} 
\includegraphics[scale=0.45]{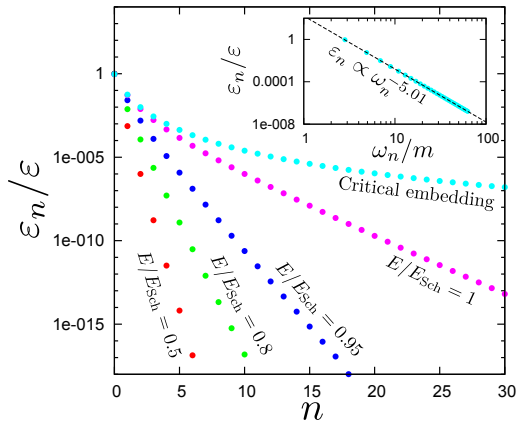}
\caption{The energy distribution for the $n$-th meson resonance.
The color of the dots follows that of the previous figure.
The inset is the log-log plot of the energy distribution for the
 critical embedding, in which 
we take the meson mass spectrum $\omega_n$ as the horizontal axis.
}
\label{finiteEenergy}
\end{figure}

Whether the Maxwell-Boltzmann law is replaced by the power-law or not can be seen clearly
when the energy deposit to each meson is measured as a function of the meson resonance mass $\omega_n$.
For the static solutions, the energy expression (\ref{energyn}) is simplified just to 
$\varepsilon_n=\omega_n^2 \bm{c}_n^2/2$.
In Fig.~\ref{finiteEenergy}, we plot the meson energy distribution as a function of the resonant meson 
level $n$. In the upper-right corner of Fig,~\ref{finiteEenergy}, we have a log-log plot and find a power-law
distribution of the energy,
\begin{eqnarray}
\varepsilon_n \propto \omega_n^{-5.01} \, .
\end{eqnarray}
Thus the power is found to be $\alpha = -5.01$.
This shows a clear result of the turbulent meson condensation. 
The scaling of the energy as a function of the meson level $n$ (which is a holographic momentum quantized)
reminds us of a Kolmogorov scaling of turbulence.

\vspace{5pt}

We conclude that for the static case with a constant electric field,
right before the quark deconfinement transition the 
meson resonances condense coherently and in a turbulent manner.

\vspace{5pt}

We have several comments. First, in our static example of the constant electric field, 
we have no condensation of the vector mesons. This is consistent: only mesons which possess
the same quantum number as the vacuum get condensed, and no further symmetry breaking
occurs at the deconfinement transition. Second, how can we see the turbulent condensation in
the meson theory? Our meson effective action (\ref{mesoneff})
includes various couplings and coefficients, which are dependent on the electric field.
Once we increase the electric field, the coefficient changes, and the mesons start
to get condensed. Mesons in a theory with a single flavor is neutral under the electric field, but
the meson can polarize and has a nonlinear dependence in $E$. In particular in this theory
there exists a one-point function of the scalar mesons for a nonzero $E$, which drives the
meson condensation.

\subsection{Relevance to the string condensation}

So far, we have considered only the scalar and the vector fields on the D7-brane.
these are massless excitations of an open string on the D7-brane. The infinite tower of 
the meson resonances which we saw above is just a Kaluza-Klein-like tower obtained 
by a decomposition in the radial holographic direction. So, intuitively, they do not directly
correspond to the long QCD strings. This seems to be unsatisfactory for our purpose,
since our power-law conjecture came from the intuitive picture of the deconfined phase
that is a condensation of long strings.

\vspace{5pt}

\begin{figure} 
\includegraphics[scale=0.5]{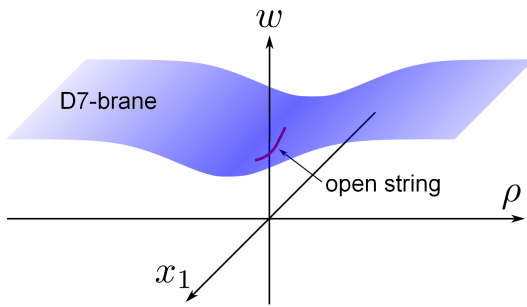}
\caption{The configuration of a fundamental string. The mesons are excited states of 
a fundamental open string hanging from the D7-brane. Lower excitations correspond to
short string, which becomes almost parallel to the worldvolume of the D7-brane.}
\label{figstringconfig}
\end{figure}

To fill this gap, we consider an open string attached to the D7-brane, see Fig.~\ref{figstringconfig}. The tension
of the open fundamental string at the tip of the D7-brane is nothing but the Regge slope,
thus is an effective QCD string tension of the theory. Now, as we turn on the electric field,
the D7-brane shape changes and the effective tension changes. In addition, since the electric field pulls the ends of the open string (quarks) directly, thus makes the effective QCD string tension reduce. We shall see that the effective QCD string goes to zero near the
deconfinement transition.

\vspace{5pt}

\begin{figure} 
\includegraphics[scale=0.7]{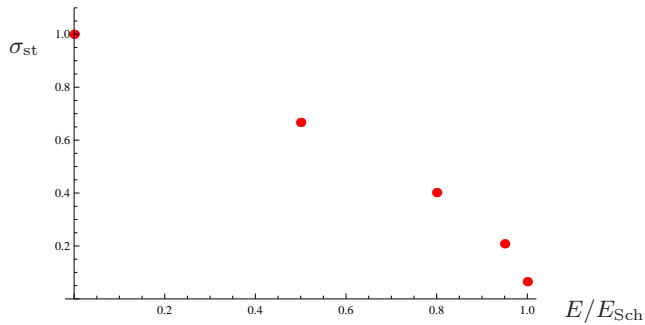}
\put(-200,100){$\sigma_{\rm st}$}
\put(10,0){$E/E_{\rm Sch}$}
\caption{The string tension $\sigma_{\rm st}(E)$ in the unit of $\sqrt{2\lambda}m^2/2\pi$.
Increase of the electric field makes the effective string tension decrease. The string tension approaches zero near the phase transition.
}
\label{figtension}
\end{figure}

The open string tension at the tip of the D7-brane is given by
\begin{eqnarray}
\sigma_{\rm st} = \frac{1}{2\pi l_s^2} \sqrt{-g_{00}g_{11}} -
 \frac{R^2}{2\pi l_s^2}E \, .
\end{eqnarray}
The open string worldsheet is put along the direction of the electric field. The
last term is the subtraction due to the electric field for giving the effective QCD string
tension. 
The metric in the Nambu-Goto action should be evaluated at the tip of the D7-brane $w(\rho=0)$, so
\begin{eqnarray}
\sigma_{\rm st} = \frac{1}{2\pi l_s^2} \left[\frac{w(\rho=0)^2}{R^2} 
-R^2 E\right] \, .
\end{eqnarray}
Since we know how the tip of the D7-brane $w(\rho=0)$ depends on the 
electric field $E$, we can numerically evaluate the tension $\sigma_{\rm st}$. The result is
shown in Fig.~\ref{figtension}. We find that the effective QCD string tension goes to zero
near the deconfinement transition. So, we conclude that not only the higher meson
resonance from the massless open string modes but also the whole string excitations
get condensed at the deconfinement transition. 

\vspace{5pt}

Generically, when the D-brane action density vanishes due to the electric field,
the electric field value is equal to the open string tension at the point of the vanishing D-brane action density \cite{Sato:2013pxa,Sato:2013hyw,Sonoda}. 
When the D-brane action density vanishes, the system is expected to undergo a phase transition to the deconfined phase, and our finding is consistent.
Furthermore, the vanishing tension of the QCD string at the deconfinement transition is
consistent with the lattice data on entropy of a heavy quark pair \cite{Hashimoto:2014fha}.

\vspace{5pt}

The string condensation shows up in string theory at finite temperature. The Hagedorn transition is related to a condensation of strings winding the temporal direction \cite{Atick:1988si, Hagedorn:1965st}. In our present case, the deformed D7-brane serves as
a probe of a horizon, and we can see the tendency to the condensation by
approaching the phase transition gradually (see \cite{Emparan:1994bt} for a discussion on how the horizon can be seen as a Hagedorn transition).

\vspace{5pt}

The vanishing of the D-brane action mimics a tachyon condensation in string theory.
When an electric field is present at the tachyon condensation, string fluid appears \cite{Bergman:2000xf,Gibbons:2000hf,Gibbons:2002tv}. The string fluid is a macroscopic
fundamental strings, and is similar to the background QCD string in spirit. 
Furthermore, when the electric field becomes critical, the original D-brane seems to
lose the property with the vanishing tension and can be of arbitrary shape; a supersymmetric example is a super tube \cite{Mateos:2001qs}. 
In this way, the vanishing D-brane action
density provides universally a condensation of macroscopic strings, and through the AdS/CFT correspondence they correspond to the QCD strings. Our example supports the generic feature of the D-branes at criticality.

\section{Meson turbulence in quenched electric fields}
\label{sec:electric_quench}

Now, we shall turn to the meson turbulence in time-dependent
set-up.
The higher meson condensation seems to be a 
sufficient cause of quark deconfinement.
In this section, this is clearly seen in a time-dependent, electric field quench
that we study below. 

\subsection{Linear theory for quenches}
\label{subsec:lineartheory}

 Before we discuss non-linear time-evolutions in the electric field quench, let us
 recall linear theory regarding quenches as perturbations. This linear analysis is also available for the quark-mass quench which
will be studied in the next section. 
By comparing analytic results of the linear analysis with the following numerical non-linear
 evolutions, we can clarify when the non-linearity will significantly affect dynamics on the brane.
Then, it turns out that the linear analysis is useful for considering initial spectrum of excitations caused by the quenches.
 
\vspace{5pt}

 The equation of motion for the fluctuations $\bm{\chi}$ is 
 \begin{equation}
  \left[\frac{\partial^2}{\partial t^2}
   - m^2 \frac{z}{1-z^2}\frac{\partial}{\partial z}
   \frac{(1-z^2)^2}{z}\frac{\partial}{\partial z}\right] \bm{\chi} = 0,
 \end{equation} 
 where $\bm{\chi}$ can describe both fluctuations of the brane position and
 the worldvolume gauge field, namely $\bm{\chi} = (W-m, a_x)$.
 Note that we can deal with the both fluctuations identically because
 mass spectra of the scalar and vector mesons 
 corresponding to the fluctuations of scalar and gauge fields respectively
 are degenerate in absence of the static electric field.

\vspace{5pt}

  The Fourier transform of a solution $\bm{\chi}$ is  
 \begin{equation}
  \hat \chi(\omega,y) = 
   \frac{1}{2\pi}\int^\infty_{-\infty}\bm{\chi}(t,z)e^{-i\omega t} dt ,
 \end{equation}
 and satisfies the differential equation 
 \begin{equation}
  y(1-y)\hat{\chi}'' - 2y \hat{\chi}' + \frac{\omega^2}{4m^2} \hat{\chi} = 0 ,
 \end{equation}
 where $y \equiv z^2$ and the prime denotes $y$-derivative.
 Imposing boundary conditions that $\bm{\chi}$ is regular at the pole ($y=1$) and satisfies 
 $\bm{\chi}(t,0)=\chi_0 (t)$ at the AdS boundary ($y=0$), we have 
 \begin{equation}
  \hat{\chi} (\omega,y) = \hat{\chi_0}(\omega)
   \frac{F(\lambda_+, \lambda_-, 2; 1-y)}
   {F(\lambda_+, \lambda_-, 2; 1)} , \label{eq:a_solution}
 \end{equation}
 where $\lambda_\pm(\omega) = (1\pm\sqrt{1+\omega^2/m^2})/2$ and 
 $\chi_0(t) = \int_{-\infty}^\infty \hat\chi_0(\omega) e^{i\omega t}d\omega$ .
 The function $\chi_0(t)$, which is assumed to be
 $\chi_0(t)=0$ for $t<0$, gives us mass or electric field quench at the
 boundary.  
Here, the latter fractional part of Eq.~(\ref{eq:a_solution}) has
 no pole in the lower half-plane of $\omega$, because it is nothing but
 a Fourier transform of a response function.
 Note that
 \begin{equation}
  F(\lambda_+, \lambda_-, 2; 1) =
   - \frac{4m^2}{\pi \omega^2} \cos\left(\frac{\pi}{2}\sqrt{1+\omega^2/m^2}\right)\, .
 \end{equation}
 Using the following formula 
 \begin{equation}
  \frac{\pi}{\cos\pi x} = - \sum_{n=0}^\infty 
   \frac{(-1)^n (2n+1)}{x^2 - (n+\frac{1}{2})^2} ,
 \end{equation}
 we can rewrite (\ref{eq:a_solution}) as 
 \begin{equation}
  \begin{aligned}
  \hat{\chi} (\omega,y) =& \hat{\chi}_0(\omega)
   \sum_{n=0}^\infty \frac{(-1)^n (2n+1)\omega^2}{\omega^2 - 4n(n+1)m^2}
   F(\lambda_+, \lambda_-, 2; 1-y) \\
   =& \hat{\chi}_0(\omega) \left[1 -  
   \sum_{n=0}^\infty \frac{(-1)^n (2n+3)\omega^2}{\omega^2 - 4(n+1)(n+2)m^2}
   \right]
   F(\lambda_+, \lambda_-, 2; 1-y) ,
  \end{aligned}
 \end{equation}
 where in the square bracket the former term is 
 the non-normalizable mode and the latter term is a summation of the normalizable
 modes with poles representing the meson masses.

\vspace{5pt}

 In the time domain, the normalizable part of the solution after the
 quench $t>0$ becomes 
 \begin{equation}
  \begin{aligned}
   \bm{\chi}(t,z) - \bm{\chi}_\mathrm{non}(t,z) =& 
   - \sum_{n=0}^\infty 
   \int_{-\infty}^\infty d\omega e^{i\omega t}
   \frac{(-1)^n (2n+3)\omega^2}{(\omega - i0^+)^2 - \omega_n^2}
   F(\lambda_+, \lambda_-, 2; 1-z^2)\hat{\chi}_0(\omega) \\
   =&  
   2\pi i \sum_{n=0}^\infty \frac{2n+3}{2} 
   \omega_n z^2 F(n+3, -n, 2; z^2) e^{i\omega_n t} \hat\chi_0 (\omega_n)
   + \text{c.c.} \\
   =& 
   2\pi i m \sum_{n=0}^\infty \sqrt{\frac{2n+3}{2}}
   \hat\chi_0 (\omega_n) e^{i\omega_n t} e_n(z) + \text{c.c.} ,
  \end{aligned}
 \end{equation}
 where we have used 
 $F(\lambda_+, \lambda_-, 2; 1-z^2)|_{\omega=\omega_n} = (-1)^n z^2
 F(n+3,-n,2,z^2)$.
 The eigen functions $e_n(z)$ and the eigen value $\omega_n$ have been defined
 by (\ref{eq:normal_mode}) and (\ref{eq:meson_mass}), respectively.
 Note that $\bm{\chi}_\mathrm{non}(t,z)$ is the non-normalizable mode.

\vspace{5pt}

 As a result, in linear response the
 quench characterized by the 
 boundary condition $\chi_0(t)$ will cause excitations with the energy spectrum  
 \begin{equation}
  \varepsilon_n \equiv \frac{1}{2}(\dot{c}_n^2 + \omega_n^2 c_n^2)
   = 4\pi^2 m \omega_n^2 \sqrt{\omega_n^2 + m^2} 
   |\hat{\chi}_0(\omega_n)|^2 ,
\label{lin_spec}
 \end{equation}
 where $\bm{\chi}(t,z) = \bm{\chi}_\mathrm{non}(t,z) + \sum_{n=0}^\infty c_n(t) e_n(z)$.


\subsection{Energy transfer to higher meson modes}

Now, we consider the electric field quench in full non-linear theory.
Assuming translational symmetry generated by $\partial_{\vec{x}}$ and
spherical symmetry of $S^3$, we can write the dynamical brane solution as
\begin{equation}
 w=W(t,\rho),\quad 2\pi \ell_s^2 L^{-2} A_x=a_x(t,\rho)\ , 
\end{equation}
where we set $\bar{w}=0$ without loss of generality.

\vspace{5pt}

\begin{figure}
\begin{center}
\includegraphics[scale=0.45]{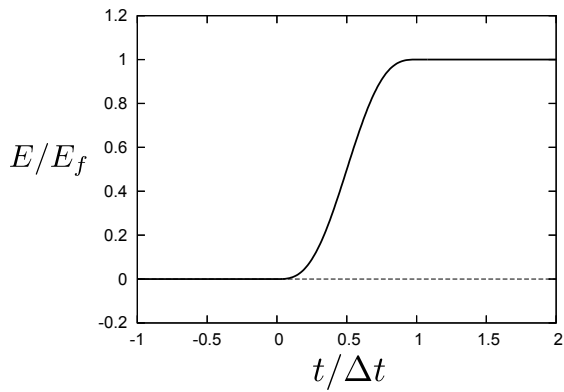}
\end{center}
\caption{
Profile of the time-dependent electric field $E(t)$.
}
 \label{Eoft}
\end{figure}

We consider a time-dependent boundary condition for $a_x$ as
\begin{equation}
 a_x|_{\rho=\infty}=-\int dt E(t)\ .
\label{axbdry}
\end{equation}
where $E(t)$ corresponds to the external electric field in the boundary
theory $\mathcal{E}(t)$ as $\mathcal{E}=(\lambda/(2\pi^2))^{1/2}E$.
Starting from the $E=0$ vacuum in the confined phase $W=m$ for $t<0$,
we turn on the electric field smoothly to reach a final value $E_f$ in the duration 
$\Delta t$. Explicitly, we choose a $C^2$ function for $E(t)$ as
\begin{equation}
E(t)=
\begin{cases}
0 & (t<0) \\
E_f[t-\frac{\Delta t}{2\pi}\sin(2\pi t/\Delta t)]/\Delta t & 
 (0 \le t \le \Delta t) \\
E_f & (t>\Delta t) 
\end{cases}
\ .
\label{Efunc}
\end{equation}
The profile of the electric field is shown in
Fig.~\ref{Eoft}.
In our previous work \cite{Hashimoto:2014yza}, we solved the brane
motion numerically imposing the boundary condition~(\ref{axbdry}).%
\footnote{In~\cite{Hashimoto:2014yza}, we used the in-going
Eddington-Finkelstein time $V\equiv t-1/r$ as the bulk time
coordinate for
the convenience of numerical calculations. At the AdS boundary
$r=\infty$, both time coordinates $V$ and $t$ mean the same boundary time.}
When the final value of the electric field $E_f$ is below the Schwinger
limit $E_\textrm{Sch}$,
at which deconfinement transition occurs, 
a pulse-like fluctuation induced by the electric field
quench at the AdS boundary propagates between the AdS boundary
($\rho=\infty$) and the pole ($\rho=0$)
on the brane worldvolume.
We found that, after several bounces, the fluctuation collapses to a naked singularity
at $\rho=0$ depending on parameters $E_f$ and $\Delta t$.
It turns out a strongly redshifted region appears near the naked singularity.
This is interpreted as an ``instability'' toward deconfinement, 
which happens, to our surprise, even when the final 
field strength is below the Schwinger limit.

\vspace{5pt}

\begin{figure}
  \centering
  \subfigure[Condensate]
  {\includegraphics[scale=0.5]{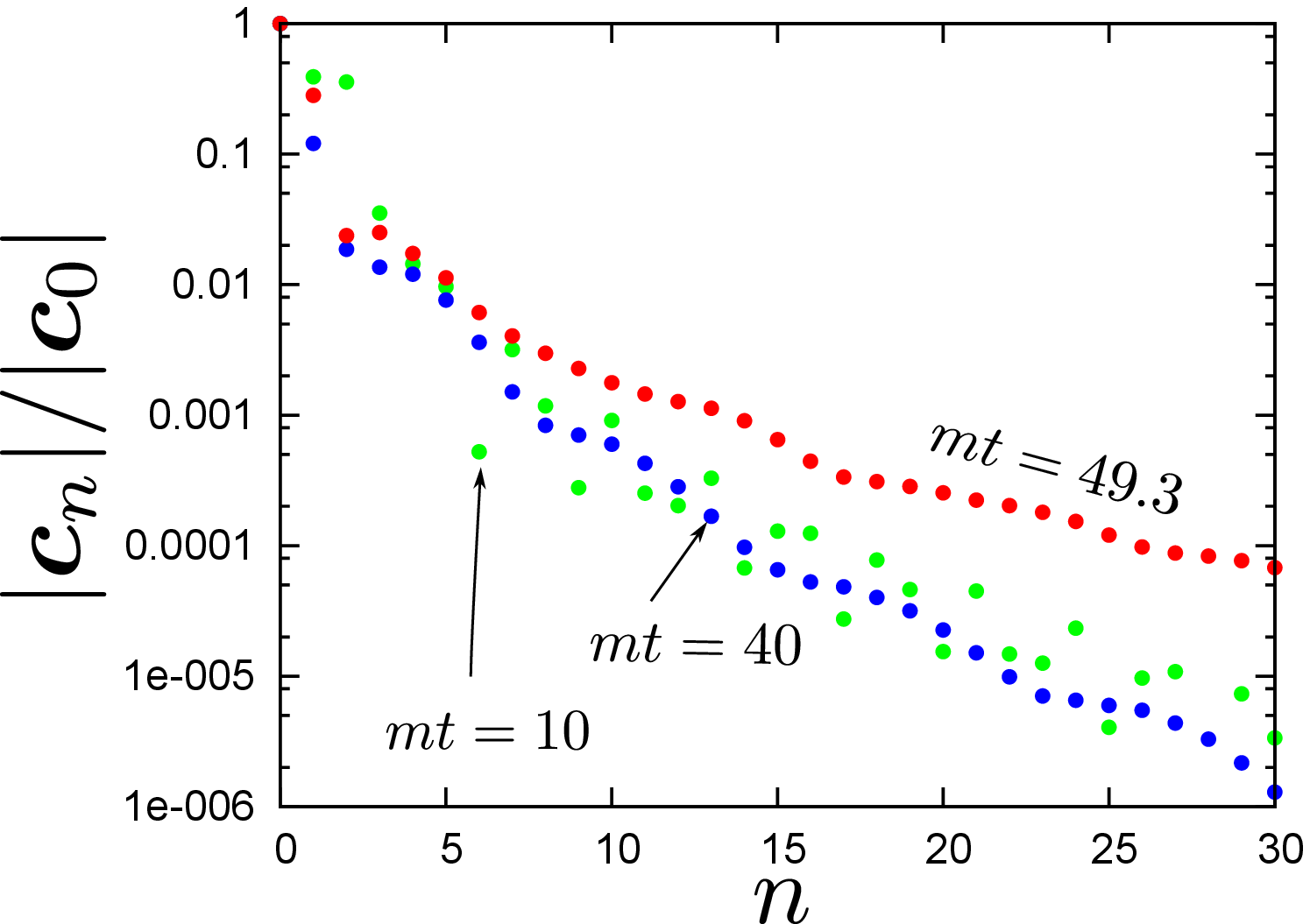}\label{nth_cond}
  }
  \subfigure[Energy spectrum]
  {\includegraphics[scale=0.5]{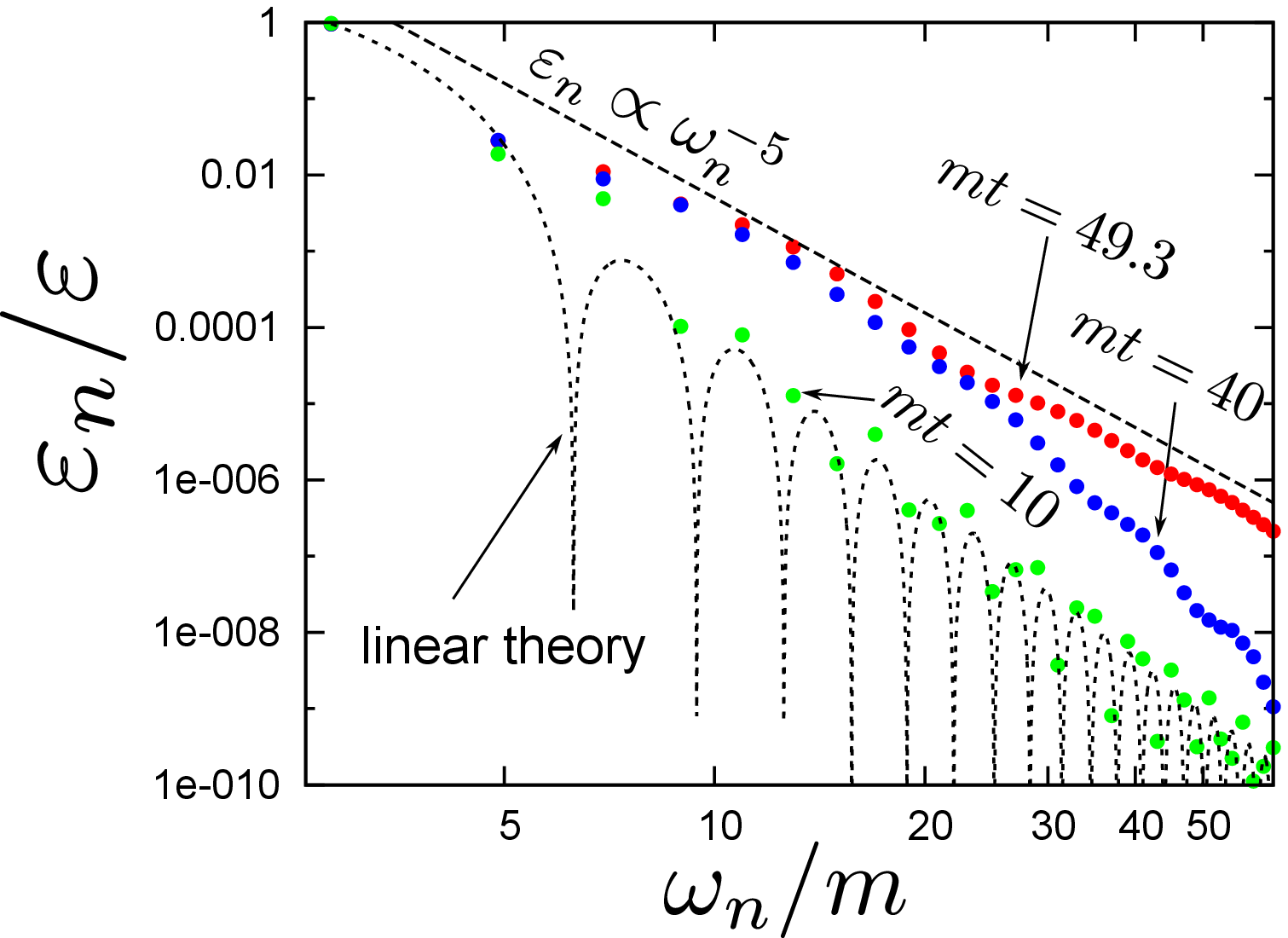} \label{nth_energy}
  }
  \caption{
(a) Condensate and (b) energy contributed by $n$-th exited meson 
induced by an electric field quench.
 Data for times $m t=10$, $40$, and $49.3$ are shown.
We set 
$E_f/E_\textrm{Sch}=0.2672$ 
and $m \Delta t=2$.
A clear (non-thermal) growth at large $n$ is found along the time
 evolution.  
The energy spectrum agrees with that of the linear theory initially and
 then seems to approach $\varepsilon_n\propto \omega_n^{-5}$ as time increases.
}
\label{nth_ce}
\end{figure}

Our finding can be considered as a probe-brane version of the weakly
turbulence similar to that of the global AdS 
spacetime~\cite{Bizon:2011gg} in which a non-linear evolution of 
a perturbed AdS spacetime causes an instability resulting in a black
hole formation. 
They studied the energy
spectrum of the perturbation 
and found that the energy is transferred from low to high frequencies as
time increases.
Following them, we also study the time evolution of energy spectrum of the
brane fluctuation in spectral analysis.
In the following, we choose a weak electric field 
($E_f/E_\textrm{Sch}=0.2672$) and a switch-on duration 
$m \Delta t=2$, in which 
sub-Schwinger-limit deconfinement is realized.
We decompose the time-dependent non-linear solutions obtained in
Ref.~\cite{Hashimoto:2014yza} into normal modes Eq.~(\ref{eq:normal_mode}) on the
supersymmetric background 
and calculate the condensate $\bm{c}_n$ and energy spectrum $\varepsilon_n$ 
of the meson resonances.
In Fig.~\ref{nth_cond}, the time evolution of the 
condensate $|\bm{c}_n|$ is shown
for several time slices $mt=10$, $40$, and $49.3$, while
the time $m t=49.3$ is just before deconfinement.
Also, that of the energy spectrum $\varepsilon_n/\varepsilon$ is shown in Fig.~\ref{nth_energy}.

\vspace{5pt}

In linear theory, from
Eq.~(\ref{lin_spec}), we obtain the energy spectrum analytically as
\begin{equation}
\begin{aligned}
 \varepsilon_n^\textrm{linear}
 =&4\pi^2 m \sqrt{\omega_n^2+m^2} |\hat{E}(\omega_n)|^2 \\
=&
\frac{32\pi^4E_f^2m\sqrt{\omega_n^2+m^2}(1-\cos\omega_n \Delta t)}
{\omega_n^4\Delta t^2(4\pi^2-\omega_n^2\Delta t^2)^2}
\ ,
\end{aligned}
\end{equation}
where we have used $\hat\chi_0(\omega) = i \hat E(\omega) / \omega$ and
$\hat{E}(\omega)=(2\pi)^{-1}\int^\infty_{-\infty} E(t)e^{-i\omega t} dt$ for the electric quench.
We show the $\varepsilon_n^\textrm{linear}$ in Fig.~\ref{nth_energy}
regarding $\omega_n$ as a continuous number for visibility.
For a large $n$, the spectrum in linear theory becomes 
$\varepsilon_n^\textrm{linear}\propto (1-\cos\omega_n \Delta t)/\omega_n^7$.
Because of the trigonometric function $\cos\omega_n \Delta t$, 
it oscillates as a function of $\omega_n$.
We find that at an early stage of the time evolution the energy
spectrum can be very well described by that of the liner theory
$\varepsilon_n^\textrm{linear}$, and then gradually evolves as the condensate and the energy are transferred to higher meson modes during the time evolution.
This tendency is similar to the static case
in which they are ``transferred'' more as electric fields are stronger. 
Eventually, the energy spectrum $\varepsilon_n/\varepsilon$ 
seems to approach $\varepsilon_n\propto \omega_n^{-5}$ just before the deconfinement, which
has been seen in the static calculation. (See Fig.~\ref{finiteEenergy}.)
It reminds us a Kolmogorov-like scaling again.
This indicates that 
higher meson condensation is universally 
related to quark confinement.

\vspace{5pt}

The observed time evolution of the distribution suggests that turbulence is taking place in the meson sector. 
This is because higher modes have smaller wave lengths 
in holographic directions, 
and the transfer of energy and momentum indicates that smaller structure are being organized 
during the time evolution toward deconfinement. 


\subsection{Condition for the turbulent meson condensation}

In Ref.~\cite{Bizon:2011gg}, 
it was suggested that the global AdS is unstable against the black hole
formation even for arbitrary small perturbations.
Here, we explore the condition for the singularity formation in the D3/D7
system with the external electric field~(\ref{Efunc}).
In our previous work~\cite{Hashimoto:2014yza}, 
we introduced the redshift factor 
$r_s(P_1,P_2)\equiv \omega(P_2)/\omega(P_1)$ associated with an out-going light ray
connecting between $P_1$ and $P_2$, 
where $P_1$ and $P_2$ are points at the initial surface and
the AdS boundary, respectively.
$\omega(P_1)$ and $\omega(P_2)$ are frequencies of
the light ray measured by natural static observes 
at each point.
Using the redshift factor, we defined the deconfinement phase in the view
of the gravity side: 
we say the point $P_2$ of the boundary is in deconfinement phase when
$r_s(P_1,P_2)$ has been very large (we use $r_s=100$ as a criterion).
We also defined the deconfinement time $t_d$ which 
is the boundary time at which $r_s=100$ is satisfied.
In Ref.~\cite{Hashimoto:2014yza}, 
we found that the redshift factor suddenly increases 
around a retarded time at which the singularity is formed.
As a result, it exceeds our criterion of the deconfinement $r=100$, and 
eventually it diverges. 
So, the deconfinement time $t_d$ can be a good indication of the
singularity formation. 
If there exists any non-zero minimum value of $E_f$ as $t_d\to \infty$, the
singularity formation does not occur for arbitrary small $E_f$.
We will seek such a threshold 
value of $E_f$ in the following.

\vspace{5pt}

\begin{figure}
\begin{center}
\includegraphics[scale=0.5]{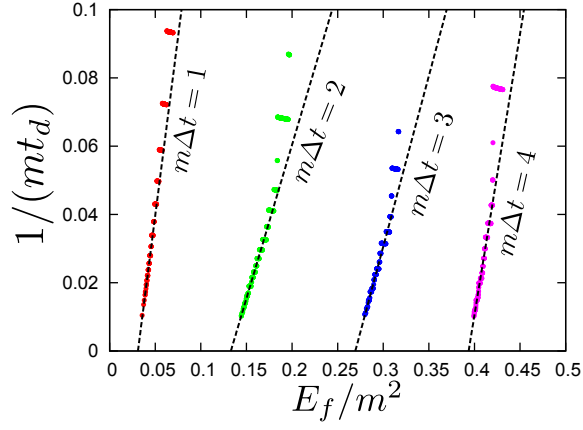}
\end{center}
\caption{
Inverse of the deconfinement time $t_d$ as a function of $E_f$.
Blue, green, red and magenta@points correspond to numerical data for 
$m \Delta t=1,2,3,4$, respectively.
Fitting line are shown by dashed lines.
The points where the dashed lines and the horizontal axes intersect
 determine $E_\infty$.
}
 \label{tdecinvE}
\end{figure}
\begin{figure}
\begin{center}
\includegraphics[scale=0.5]{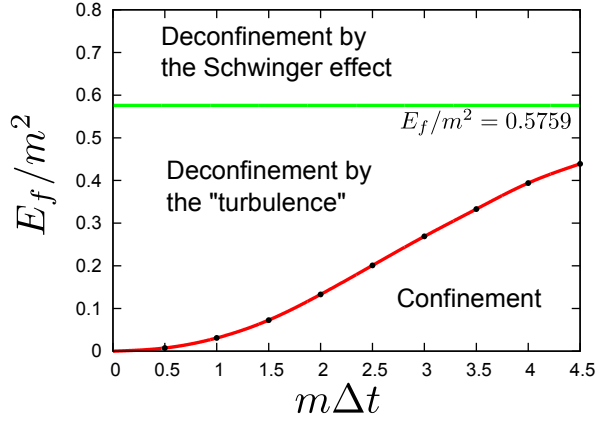}
\end{center}
\caption{
Dynamical phase diagram for the electric field quench.
}
\label{dynph}
\end{figure}

In Fig.~\ref{tdecinvE}, we plot the inverse of the deconfinement time
$1/(mt_d)$. Note that the deconfinement time is a discrete function as
mentioned in Ref.~\cite{Hashimoto:2014yza}.
We can find that $1/(mt_d)$ seems to approach the horizontal axis  
and touch there at non-zero $E_f$ by extrapolating.
Thus, the deconfinement time behaves as 
$t_d \sim 1/(E_f-E_\infty)$ as $E_f \to E_\infty\neq 0$.
This indicates that the singularity
formation does not occur for sufficiently small $E_f$.
We fit the numerical data by linear functions and show them by dashed lines. From 
the fitting line, we estimate the lower
bound of the singularity formation $E_\infty$, at which the
deconfinement time $t_d$ will be infinite. 
In Fig.~\ref{dynph}, we plot $E_\infty$ 
for several $\Delta t$.
Our numerical data are shown by black points.
They are interpolated by a spline curve passing through
the origin. This figure can be regarded as ``dynamical phase diagram''
for the electric field quench in supersymmetric QCD:
(i) Below the red curve, we cannot observe the singularity formation.
Thus, the system is in the confinement phase in this parameter region.
(ii) Above the red curve and below the green curve, we find the naked
singularity formation.
Since the singularity formation implies the deconfinement, 
this region is regarded as a transient deconfinement phase.
(iii) Above the green curve, we find the effective horizon on the D7-brane and
the system settles down to a stationary phase eventually. 
This region can be regarded as the deconfinement phase induced by the Schwinger effect. 



\section{Exploration of the essence of the turbulence}
\label{sec:mass_quench}

In the previous section, we have mentioned the turbulent meson
condensation applying the quenched electric field.
Now, the following questions arise: 
What is essential to the turbulent meson condensation?
Does the worldvolume gauge field play an important role?
If the electric field is present, the system has many elements
suspected as a cause of the turbulence; 
two components of the fluctuations corresponding to the scalar and
vector mesons, the mode mixing between those components in the existence
of the static electric field (Stark effect), and so on.
In this section, to answer these questions, 
we study time evolution of the D7-brane in 
the simplest set up: quark-mass quench.

\subsection{Set up}

\begin{figure}
\begin{center}
\includegraphics[scale=0.4]{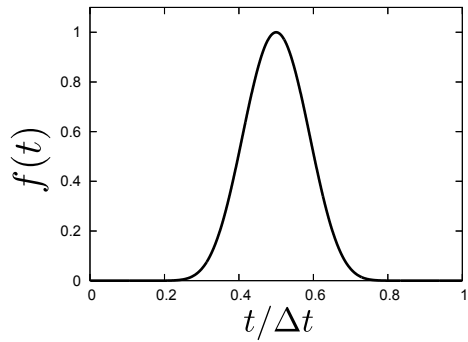}
\end{center}
\caption{
Profile of the function $f(t)$ for $m=1$.
}
 \label{Cinf}
\end{figure}

We set the gauge potential to be zero and 
solve the dynamics of the D7-brane with time-dependent boundary condition as
\begin{equation}
 W(t,\rho=\infty)=\mu(t)\ ,
\end{equation}
where, $\mu(t)$ corresponds to the time-dependent quark mass.
In this case only the brane scalar fluctuations will be excited.
We explicitly consider the following quark mass quench: 
\begin{equation}
 \mu(t)=m+\delta m \, f(t)\ ,\quad
 f(t)\equiv
\begin{cases}
\exp\left[\frac{4}{m}\left(
\frac{1}{t-\Delta t}-\frac{1}{t}+\frac{4}{\Delta t}
\right)\right] &(0<t<\Delta t) \\
0 &(\textrm{else})\ .
\end{cases}
\end{equation}
The function $f(t)$ is a compactly supported $C^\infty$ function whose
maximum value is unity. The profile of the function is shown in
Fig.~\ref{Cinf}.
Therefore, the quench is characterized by two parameters $\Delta t$ and
$\delta m$, which means the quark mass increases by $\delta m$ and then
return to original value between the duration $\Delta t$.

\vspace{5pt}

Before the quench $t<0$, we assume that the brane is static, that is,
$W(t,\rho)=m$. For $t>0$, the brane moves in AdS$_5\times S^5$ because of the
time dependence of the boundary condition.
We solve the time evolution numerically.
We follow the numerical method developed in Refs.~\cite{Ishii:2014paa,Hashimoto:2014yza}.

\subsection{Formation of a naked singularity}

\begin{figure}
  \centering
  \subfigure[Ricci scalar]
  {\includegraphics[scale=0.45]{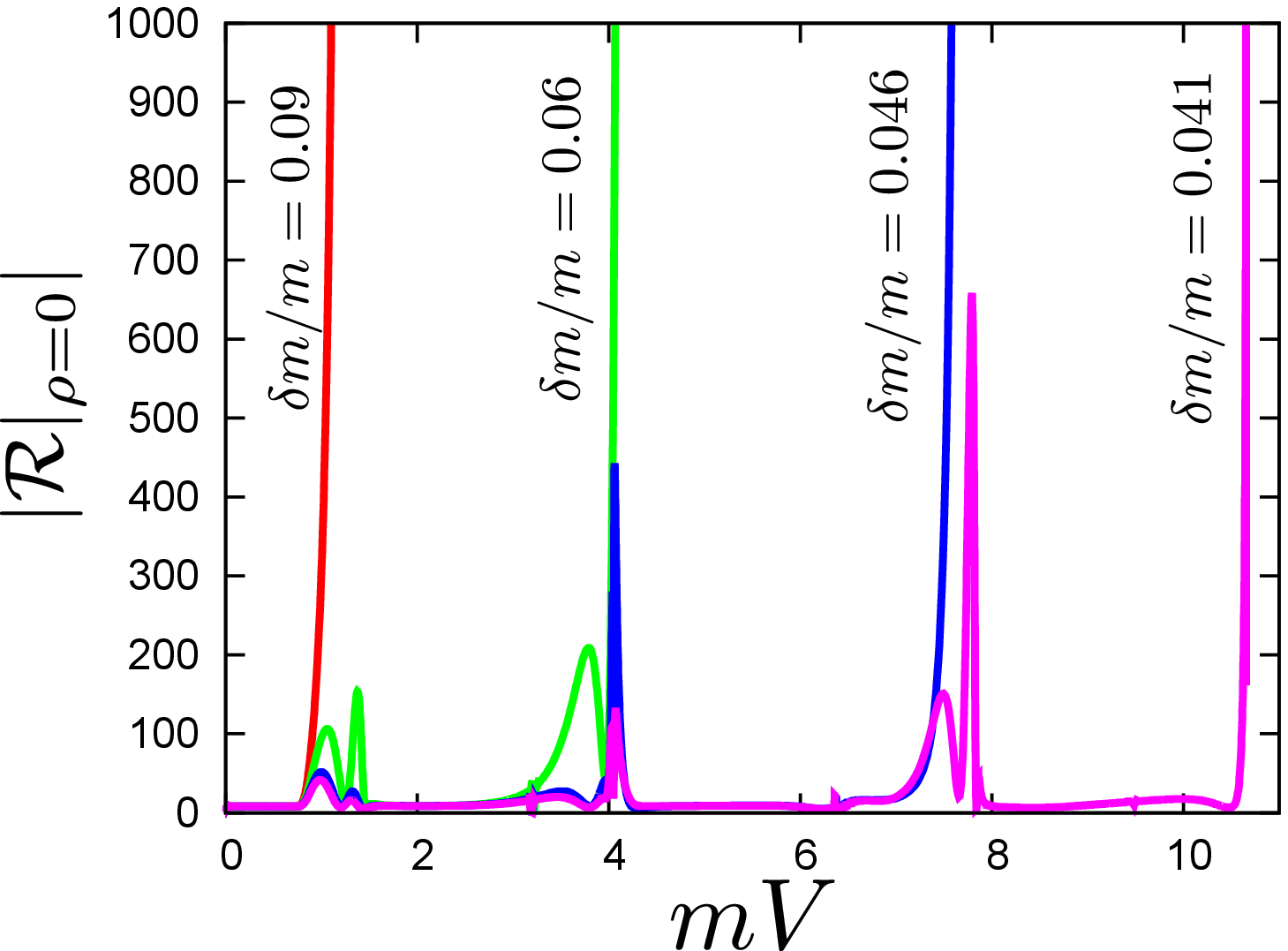} \label{RS}
  }
  \subfigure[Redshift factor]
  {\includegraphics[scale=0.45]{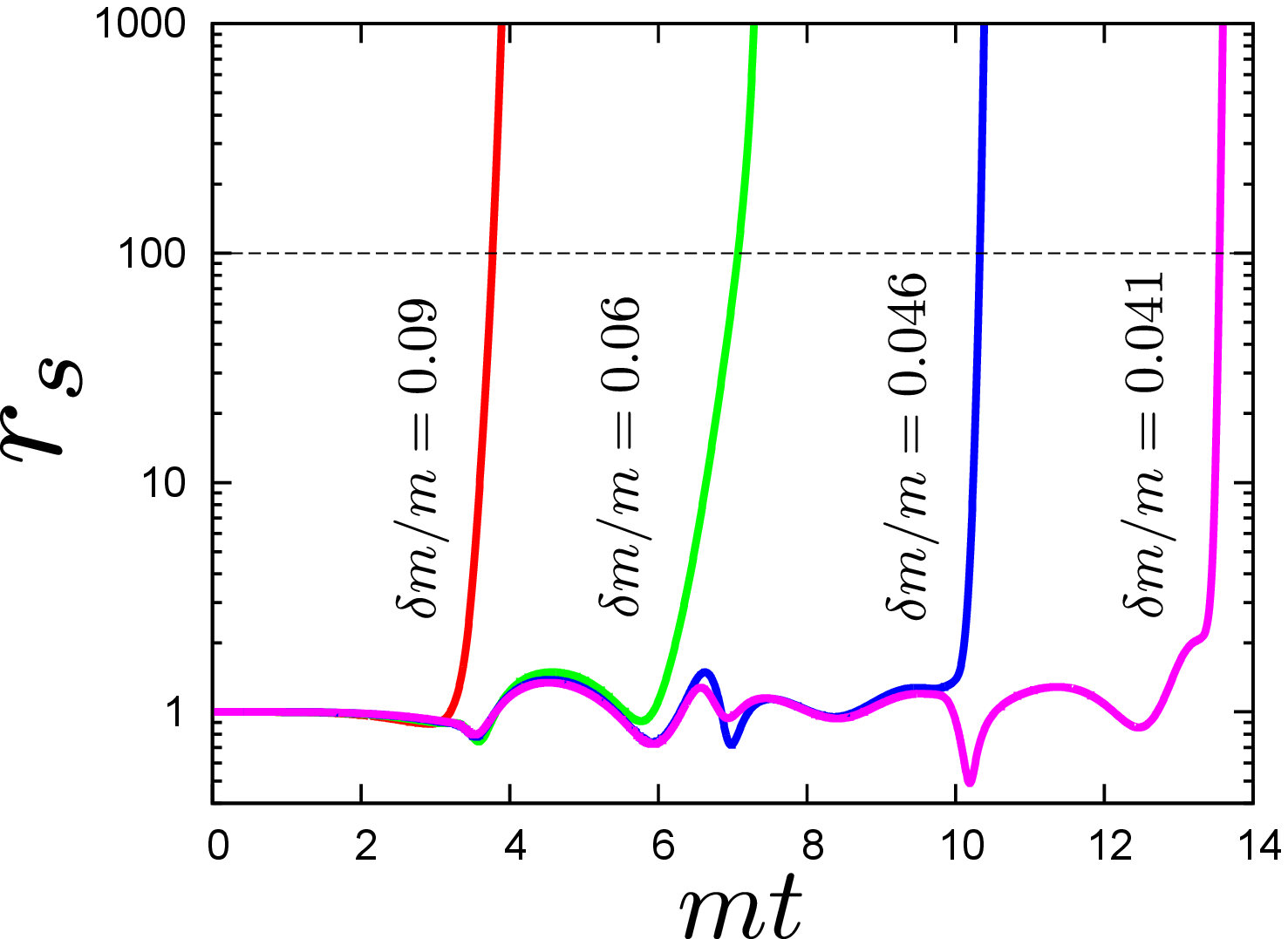} \label{redshift}
  }
  \caption{
(a) Ricci scalar estimated at the pole $\rho=0$ against $V\equiv t-1/r$.
(b) Redshift factor against the boundary time $t$
for $m\Delta t=4$.
In these figures, 
we fixed the time scale of the mass quench as $m\Delta t=4$ and varied 
its amplitude as 
$\delta m/m=0.09$, $0.06$, $0.046$ and $0.041$.
}
\end{figure}

As the result of the numerical calculation, 
we found the similar behavior as the electric field quench:
A pulse-like fluctuation induced by the mass
quench propagates between the AdS boundary and the pole.
After several bounces, the fluctuation collapses to a naked singularity
depending on parameters $\delta m$ and $\Delta t$.
To see the singularity formation in the mass quench, 
we evaluate the Ricci scalar $\mathcal{R}$ with respect to the brane induced metric
$h_{ab}$.
In Fig.~\ref{RS}, we show the Ricci scalar monitored at 
$\rho=0$ as a function of $V\equiv t-1/r$.
We can see that pulses are localized in several time intervals.
This is because the pulse-like fluctuation induced by the mass quench 
propagates between the AdS boundary and the pole.
When the pulse arrive at the pole, the brane is strongly bended and the scalar
curvature has large value.
We can find that the Ricci scalar diverges
at several places depending
on the amplitude of the mass quench.
The pulse is getting sharp as time increases and, eventually, it
collapse to a naked singularity.
For example, for $\delta m/m=0.046$, the scalar curvature diverges when
the pulse comes to the pole for the third time.
This means that, even if there is no gauge field on the D7-brane, 
a naked singularity can be created by the dynamical process.

\subsection{Deconfinement time}

\begin{figure}
  \centering
  \subfigure[$m \Delta t=2$]
  {\includegraphics[scale=0.45]{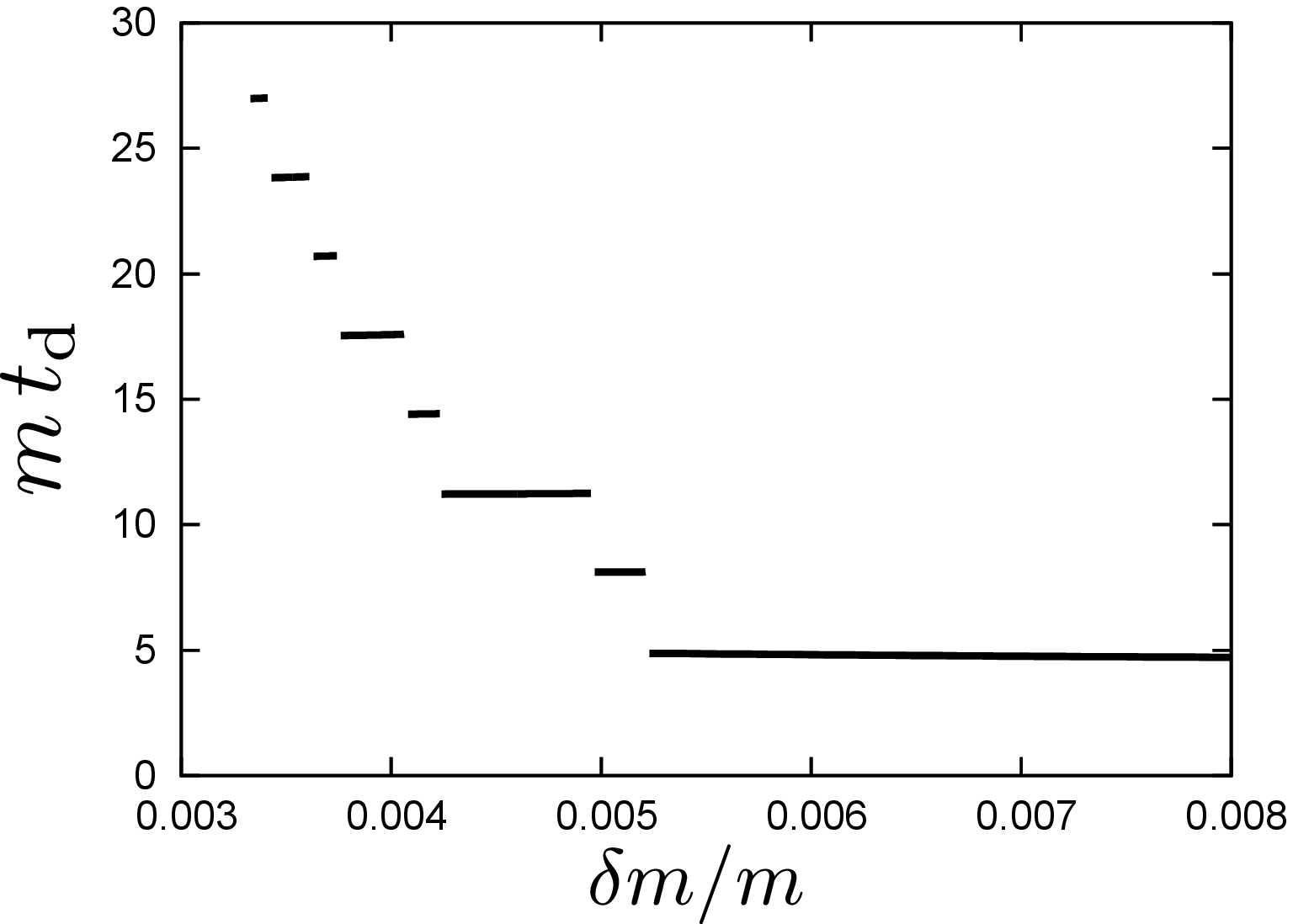}
  }
  \subfigure[$m \Delta t=4$]
  {\includegraphics[scale=0.45]{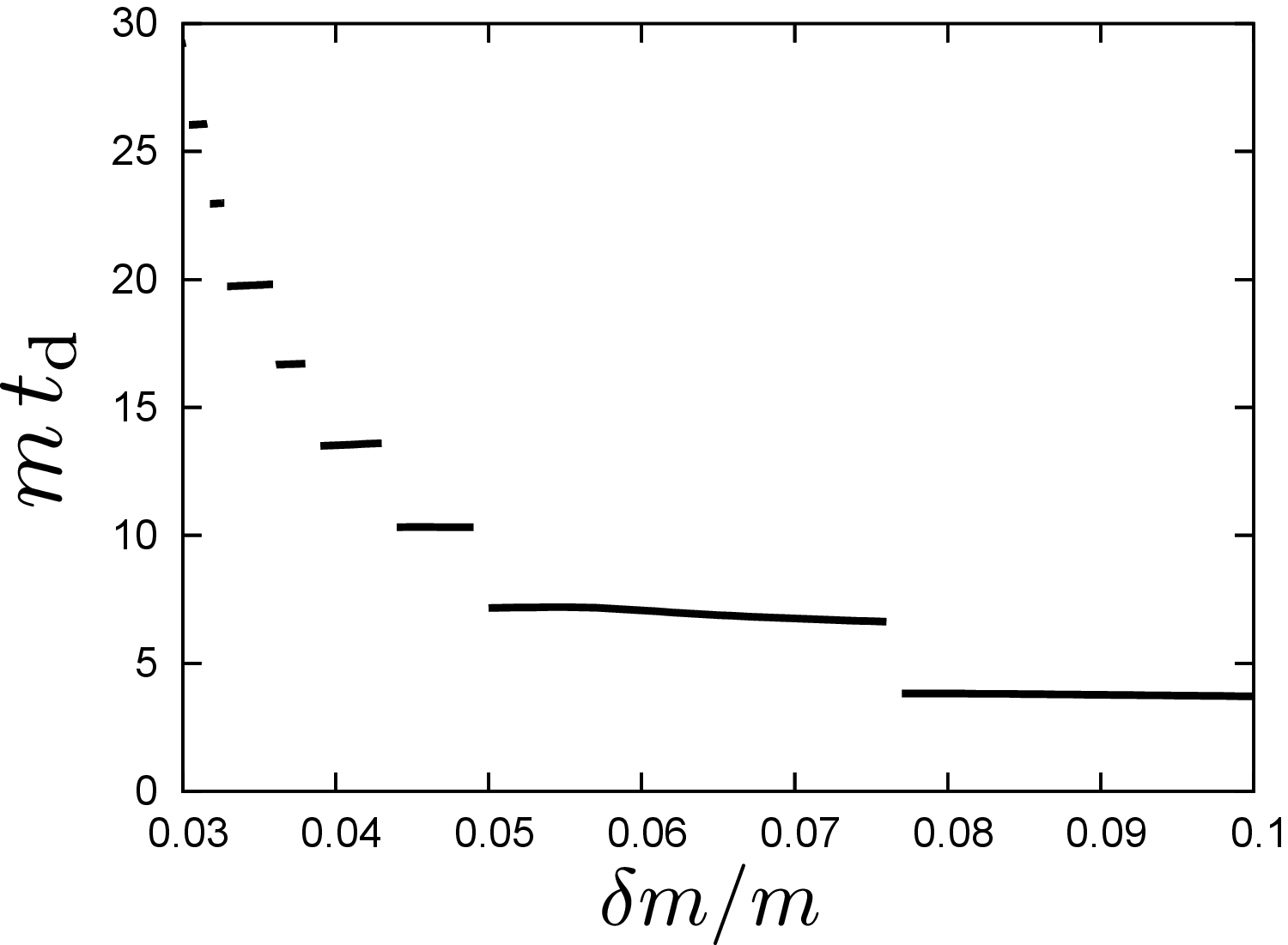} 
  }
  \caption{
Deconfinement times $t_d$ against $\delta m$ for $m\Delta t=2,4$.
They are given by discrete functions of
$\delta m$.
}
\label{tdec}
\end{figure}

Following Ref.~\cite{Hashimoto:2014yza}, we study the deconfinement time.
In Fig.~\ref{redshift}, we show the redshift factor as the function of the
boundary time $t$. The redshift factor suddenly increases 
around the retarded time at which the singularity is formed.
As a result, it exceeds the criterion of the deconfinement $r_s=100$ and 
eventually, it diverges. 
In Fig.~\ref{tdec}, the deconfinement time $t_d$ is shown as a function
of $\delta m$. We can see that the $t_d$ is a discrete function of
$\delta m$ like as the electric field quench. 
This is because the number of bounces needed for the
singularity formation depends on the $\delta m$ as shown in the previous
subsection.

\vspace{5pt}

\begin{figure}
\begin{center}
\includegraphics[scale=0.45]{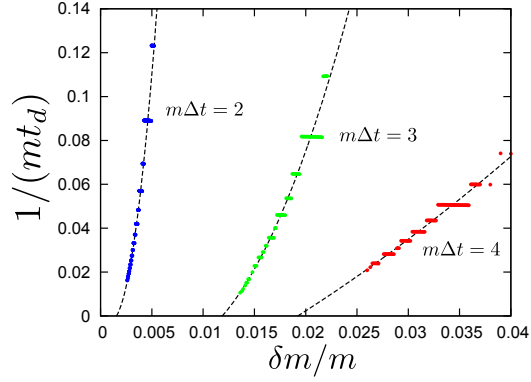}
\end{center}
\caption{
Reciprocal for the deconfinement time $t_d$ as a function of $\delta m$.
Blue, green and red points correspond to numerical data for 
$m \Delta t=2,3,4$, respectively.
Fitting curves are shown by dashed curves.
}
 \label{tdecinv}
\end{figure}

Now, we check if the turbulent behavior can occur for arbitrary small
$\delta m$. Following the case of the electric field quench, we focus on
the inverse of the deconfinement time.
In Fig.~\ref{tdecinv},
we plot $1/(mt_d)$ against $\delta m/m$ for $m\Delta t=2,3,4$.
Our numerical data are shown by the points in the figure.
We fit the numerical data by a second order polynomial 
$a \delta m^2 + b \delta m +c$. The fitting curves are shown by dashed
curves in the figure.
The dashed curves intersect with the horizontal axis.
Thus, the deconfinement time behaves as 
$t_d \sim 1/(\delta m-\delta m_\infty)$ as $\delta m \to \delta m_\infty\neq 0$.
The critical value of $\delta m$ at which the deconfinement time will be
infinite can be estimated as 
$\delta m_\infty /m=0.0016, 0.0119, 0.0192$ for $m\Delta t=2, 3, 4$, respectively.
We find that they are non-zero value. This indicates that the singularity
formation does not occur for sufficiently small perturbations.

\vspace{5pt}

\subsection{Meson turbulence in mass quench}

\begin{figure}
\begin{center}
\includegraphics[scale=0.5]{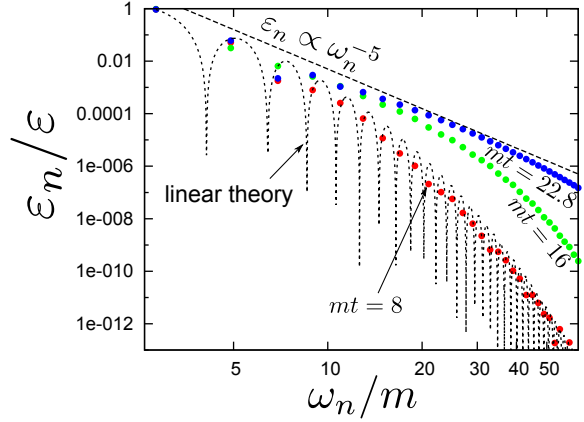}
\end{center}
\caption{
Energy spectrum for several time slices, $mt=8, 16, 22.8$.
Parameters as $\delta m/m=0.031$ and $m\Delta t=4$. It is given by the
 linear theory at first, and then 
seems to approach $\varepsilon_n\propto \omega_n^{-5}$ as time increases.
}
 \label{nth_e}
\end{figure}

In order to perform spectral analysis, 
we decompose the non-linear solution 
$w(t,z)=W(t,z)-m$ obtained numerically into the meson eigen modes
$e_n(z)$ around the supersymmetric background.
Setting parameters as $\delta m/m=0.031$ and $m\Delta t=4$, 
in Fig.~\ref{nth_e}, we plot the spectrum for several time slices, $mt=8$, $16$, and $22.8$, while
the time $m t=22.8$ is just before deconfinement.
For the mass quench, the energy spectrum in linear theory computed from
Eq.~(\ref{lin_spec}) is given by 
\begin{equation}
 \varepsilon_n^\text{linear} = 4\pi^2 m \omega_n^2 
  \sqrt{\omega^2_n + m^2} \delta m^2 |\hat{f}(\omega_n)|^2 , 
\end{equation}
where we have used $\hat\chi_0 (\omega) = \delta m \hat{f}(\omega)$ and
$\hat{f}(\omega)=(2\pi)^{-1}\int^\infty_{-\infty} f(t)e^{-i\omega t} dt$.
We also show the spectrum in the figure
regarding $\omega_n$ as a continuous number.
We can see that the energy spectrum $\varepsilon_n/\varepsilon$ is given by the linear theory at first, and
then evolving with energy flow from low to high
frequency modes because of non-linearity.
This means that, even though the worldvolume gauge field
dose not exist, the turbulent meson condensation can arise from 
the brane fluctuations $w(t,z)$ only.
We also see that 
the energy spectrum 
seems to approach $\varepsilon_n\propto \omega_n^{-5}$ just before the
deconfinement in the quark mass quench as well as the electric field
quench.
This implies that Kolmogorov-like scaling 
$\varepsilon_n\propto \omega_n^{-5}$ is universal for the deconfinement in
$\mathcal{N}=2$ supersymmetric QCD.

\subsection{Essence of the turbulence}

The results shown in the previous and this sections lead us the following conclusions.
The worldvolume gauge field
or the quench induced by the electric field is not indispensable for the
probe-brane turbulence.
In addition, although the existence of the static electric filed causes the mass
shift of the spectrum and also the mode mixing between the scalar and the vector
mesons, it does not spoil the turbulence. 

\vspace{5pt}

Since the initial energy spectrum excited by the quenches can be determined by
the linear theory, non-linearity has not been so significant to excite
the brane.
It means that perturbative (not infinitesimal but finite) inputs are enough for the turbulence on the brane.
The non-linearity plays an important role in the evolution with energy
transfer to higher meson modes after the brane is excited.

\vspace{5pt}

Here, we point out the relation with the present result and 
the ``weak turbulence theory'' developed in 
plasma physics \cite{Coppi69,Sagdeev69}. 
When the electric field is switched on or the quark mass changes, 
the meson modes are excited, and their dynamics is described by the
time evolution of mode coefficients $c_n(t)$. 
At lowest order, 
the meson modes obey a kinetic equation 
\begin{eqnarray}
\frac{d \tilde{c}_n(t)}{dt}
=\sum_{n,n',n''}W_{nn'n''} \tilde{c}_{n'}(t) \tilde{c}_{n''}(t)+
\ldots,
\label{eq:weakturbulence}
\end{eqnarray}
where the coupling constant $W_{nn'n''}$
represents 3-meson interaction that is generically
present in the effective meson  
Lagrangian. 
One can obtain the coupling 
by expanding the D7-brane action  \cite{Kruczenski:2003be},
and the kinetic equation can be obtained
as an equation of motion derived from the effective Lagrangian
(after a suitable redefinition of the meson fields\footnote{Free mesons oscillate with a trivial phase factor with the mass $\omega_n$, since the free equation of motion for
the homogeneous meson is $(\partial_t^2+\omega_n^2) c_n(t)=0$. To get rid of this phase factor, we need to redefine $c_n(t) \equiv b_n(t) \exp[i \omega_n t]$. Then this free $b_n(t)$ obeys the equation $(\partial_t^2 + 2 i \omega_n \partial_t)b_n(t)=0$, so we further redefine $(\partial_t + 2i \omega_n)b_n(t)\equiv \tilde{c}_n(t)$ such that the free equation of motion is $\partial_t \tilde{c}_n(t)=0$. The weak turbulence equation (\ref{eq:weakturbulence}) uses this eigenbasis.}). 
Energy conservation 
is weakly enforced
in $W_{nn'n''}$, {\it i.e.}, $\omega_n\sim \omega_{n'}+\omega_{n''}$,
where the excess energy is absorbed by 
other degrees of freedom. 
Equation (\ref{eq:weakturbulence}) is nothing but the 
canonical model for weak turbulence \cite{Coppi69}
and describes how the excitation spreads out 
among the meson modes. 
Before the quenches,
the mesons are not excited ($c_n(-\infty)=0$). 
Cranking up the field, the modes are excited 
as the linear theory predicts (\ref{lin_spec}), 
while the coupling 
constant $W_{nn'n''}(t)$ is also modified since the 
electric field changes the properties of the mesons. 
If the final field strength is small, 
the change of $W_{nn'n''}$ is not significant. 
The time evolution of the 
meson modes and how the energy is transfered to 
higher energies are described by the 
 kinetic equation (\ref{eq:weakturbulence}). 
The power law distribution 
we find in Figs.~\ref{nth_energy} and \ref{nth_e}
implies a scaling behavior for the coupling constant. 
It is an interesting future problem to 
explain this from the generic properties of the
effective meson theory, which 
will prove the universality of the 
meson turbulence phenomenon. 

\vspace{5pt}

Although our numerical calculations have been performed only for several variations of the quenches characterized
by specific functions, a functional form of the quench can always be mapped to an initial energy spectrum by the
linear theory. Thus, we need to discuss the inevitability of the meson turbulence by the initial spectrum, not by the
quench function itself. This implies that anything which can excite the brane will be a trigger of the turbulence.
Once an initial spectrum is given, the energy distribution evolves according to the
kinetic equation of weak turbulence (\ref{eq:weakturbulence}), and under a certain kind of
scaling property for the coupling $W$ the system is expected generically to flow to a turbulence-like power law.
Indeed, if we change the mass of the black hole in
the bulk by the method in Ref.~\cite{Ishii:2014paa}, namely temperature quench, we can confirm occurrence of the meson turbulence.
It is expected that the turbulence can universally occur for wider systems of fluctuations
of the probe branes.


\section{Summary}

In this paper, we have demonstrated in ${\cal N}=2$ supersymmetric QCD with large $N_c$ and strong coupling limit, that a meson turbulence is universal at quark deconfinement, by using the AdS/CFT correspondence. The ``meson turbulence'' is defined as a power-law distribution of the energy $\varepsilon_n$ for the $n$-th meson resonance with mass $\omega_n$, 
\begin{eqnarray}
\varepsilon_n \propto (\omega_n)^\alpha
\label{result}
\end{eqnarray}
with a certain negative parameter $\alpha$ which is unique to the theory.
We have studied various ways to produce the quark deconfinement; (1) static electric field, (2) electric field quench, and (3) mass quench. For (2) and (3), we can choose various sets of external parameters such as the final value of the electric field and the duration of the change of the electric field. Analyses in all the cases lead to a universal power 
\begin{eqnarray}
\alpha = -5
\end{eqnarray}
for the ${\cal N}=2$ supersymmetric QCD. This surprising universality of the turbulent meson condensation suggests that any QCD-like theory may have its own $\alpha$ at
quark deconfinement caused by any means.

\vspace{5pt}

As we have described in the introduction, normally one expects that the energy storage for mass $\omega_n$ meson should be $\sim \exp[-\omega_n/T]$ where $T$ is a finite temperature. QCD-string theory suggests that this exponentially suppressed behavior turns to a power-law near the critical point of the phase transition. We have confirmed this change to the power-law in our AdS/CFT calculation of the quarks deconfinement.

\vspace{5pt}

Our finding (\ref{result}) is a kind of weak turbulence (energy cascade from low momenta to higher momenta), if we see the meson mass $\omega_n$ as a momentum.
Indeed, the meson mass is a momentum in the holographic direction in AdS/CFT correspondence. So our energy distribution is a turbulence in holographic space, so may be called as a ``holographic turbulence''.

\vspace{5pt}

In the course of investigating the conditions for the quark deconfinement, we found an interesting ``dynamical phase diagram'' of the ${\cal N}=2$ supersymmetric QCD, see
Fig.~\ref{dynph}.
The standard case with static electric field corresponds to the right edge of the diagram (that is $\Delta t=\infty$, {\it i.e.} adiabatic limit), where there exists a clear critical value $E_{\rm Schwinger}$ (the green line) below which we have the confined phase. However in the time-dependent case we found that the deconfinement line is lowered significantly. In particular, when the quench becomes more abrupt, the final value of the electric field can be smaller for the deconfinement to occur. We worked at the supersymmetric QCD at large $N_c$ limit and the theory is different from QCD, but our result is quite suggestive to heavy ion experiments at which we know there are time-dependent electric field --- even though the magnitude of the electric field is small compared to the QCD scale, if the time-dependence is abrupt enough, the
electric field can help the deconfinement to occur.

\vspace{5pt}

However, the large $N_c$ limit would be very important for our meson turbulence.
The higher resonant states in reality have broader widths, and only at the large $N_c$ limit we can clearly see the resonances. In fact, the nearly-free QCD-string picture is
a good picture at large $N_c$, as the sting interaction such as joining-splitting and reconnection is measures in the unit of $1/N_c$. In reality with $N_c=3$, longer strings
are not favored and strings tend to break. Nevertheless, we hope that the analyses at large $N_c$ can provide at least an alternative and interesting picture of quark confinement/deconfinement.


\section*{Acknowledgment} 
K.~H.~would like to thank A.~Buchel, H.~Fukaya, D.-K.~Hong, N.~Iqbal, K.-Y.~Kim, J.~Maldacena, S.~Sugimoto, S.~Terashima, S.~Yamaguchi and P.~Yi for valuable discussions, and APCTP focus week program for its hospitality.  
This research was
partially supported by the RIKEN iTHES project.


\end{document}